# Plagiarism Detection: Keeping Check on Misuse of Intellectual Property

Iti Mathur
Department of Computer Science
Apaji Institute
Banasthali University
Rajasthan, India
mathur_iti@rediffmail.com

Nisheeth Joshi
Department of Computer Science
Apaji Institute
Banasthali University
Rajasthan, India
nisheeth.joshi@rediffmail.com

## ABSTRACT

*Today, Plagiarism has become a menace. Every journal editor or conference organizers has to deal with this problem. Simply Copying or rephrasing of text without giving due credit to the original author has become more common. This is considered to be an Intellectual Property Theft. We are developing a Plagiarism Detection Tool which would deal with this problem. In this paper we discuss the common tools available to detect plagiarism and their short comings and the advantages of our tool over these tools.*

## 1. INTRODUCTION

Human desire is perhaps the most intriguing and dangerous of all evils. As per the Bible[1], the form of destructive desire is termed as lust. In the modern times, due to this lust, we have witnessed great wars, economic slowdowns, thefts etc. In academics too, we are not left out of this lust. In order to gain fame quickly, people try to steal the assets of fellow academicians. The assets here are not the tangible assets, but the ideas of an academician and the research work that he has done. This work is also termed as Intellectual Property of the person. Now, a question comes into a mind. Can one actually copy one's ideas and work? The answer is yes, these days; this is a very common phenomenon. Almost all conferences and journals have to deal with this problem and almost all the journals have a policy against this kind of theft. They call is anti-plagiarism policy. So, what actually is plagiarism? Some of the people think that it is simply copying somebody's work or borrowing an idea. But plagiarism is a more grave crime then simply copying or borrowing. As per the Merriam Webster Online Dictionary [2], plagiarism is:

1. To steel and pass someone else's idea or work to pass as one's own (without giving due credit).
2. To commit literary theft.

Plagiarism word came from Latin word 'plagiarius' which means kidnapper and from Greek word 'plagion' which mean 'kidnapping'. One might ask how can taking someone else's idea without giving credit be a theft. According to World Intellectual Properties Organization [3]: "inventions, ideas and words too are the intellectual properties and are protected on similar grounds as that of an original invention. The punishment of stealing the ideas and work is the same as stealing the inventions."

Menace of plagiarism dates back to 1402 where a Spanish artist copied the painting of another artist [4]. Since then during the entire Renaissance Age and in the Modern Times, cases of plagiarism have seen an invariable explosion. The most recent and popular case is Jayson Blair Scandal, which led two of the top editors of New Your Times cost their jobs.

In India, we have Intellectual Property Rights Office, which takes care of Intellectual Property Rights. Although, Unlike United States and European Union, there is no registered case of against plagiarism, but, things look optimistic, after the amendments of 2001 in the act. In this paper we would deal with the issue of plagiarism detection and not of avoidance, one can always refer to Internet and literature to study about plagiarism avoidance.

## 2. TYPES OF PLAGIARISM

Plagiarism can broadly be classified into two categories: I. Source not cited and II. Source cited but still plagiarized. Following subsections gives a brief description of each of these categories.

## 2.1 Source Not Cited
In this type of plagiarism the writer copies data from some source and does not cite the text from where the matter has been copied. This category has various types of plagiarism, they are:

### 2.1.1 Ghost Writer
Writer turns in somebody else's work as if it is his own i.e the write makes a ditto copy of somebody else's work, word to word as if the matter written was his own.

### 2.1.2 Photocopy
Writer directly copies a significant amount of text form a single source, without any alterations. This type of plagiarism is committed by first time or novice copiers.

### 2.1.3 Potluck Paper
Writer tries to hide plagiarism by copying text from several sources and applying a little rephrasing of the sentences to make them fit together while retaining most of the original phrases.

### 2.1.4 Poor Disguise
In this type, Writer retains the essential contents form source text, while altering papers appearance and keywords and phrases.

### 2.1.5 Labour of Laziness
Writer tries to paraphrase the text from various different sources and makes the text from these papers fit together, instead of putting in some time and effort for writing the original content.

### 2.1.6 Self Stealer
Writer in this category, tries to copy his own published work and violating the originality policy of most of the academic bodies. Such type of plagiarism is often referred to as Self Plagiarism,

## 2.2 Source cited but still plagiarized
In this category, writer cites the original work but still has committed plagiarism by copying or rephrasing text form the original text. Various types of plagiarism in this category are

### 2.2.1 Forgotten Footnote
In this category, writer mentions authors name as source, but does not provide any reference or footnote for the referenced material.

### 2.2.2 Misinformed
Here, author gives wrong references and links, which makes information inaccessible, making it almost impossible to find the source.

### 2.2.3 Too Perfect Paraphrase
Here, author cites the reference but does not provide the proper way of writing down the required text i.e. quotation marks are not provided over the copied text, as if attributing basic idea to the source. Here, author is falsely claiming original work of others.

### 2.2.4 Resourceful Citer
Here, author properly cites references, paraphrasing and using quotations appropriately. Here plagiarized text is merged with original work making it difficult to differentiate because it looks like a well researched document.

### 2.2.5 Perfect Crime
Here, author poorly cites some work and ignores other. Making it look like original work.

## 2.3 Some other forms of Plagiarism
Besides these two broad categories, there are some other forms of plagiarism, they are:

### 2.3.1 Minimal Plagiarism
This is perhaps the most common form of plagiarism. Here, author uses someone else's ides, concepts, thoughts and writes it in different words and flow. Although, most people and professional bodies do not regard it as plagiarism, this is, at times referred to as stealing other people's ideas. This form of plagiarism requires lot of paraphrasing.

### 2.3.2 Partial Plagiarism
Here, text is copied from two to three different sources which makes rephrasing and synonym in the text appear as vague, then it is known as partial plagiarism. Here, the author applies some original thoughts, but lack of knowledge about the subject makes it partial plagiarism.

### 2.3.3 Full Plagiarism
Here, the entire text is copied without even changing the idea or rephrasing text or words. This is most common form of plagiarism among students who wish to complete their assignment as quickly as possible without even applying their brains.

# 3. RELATED WORK

There are various plagiarism detection tools available in the market. Some of the popular ones are:

## 3.1 EduTie
EduTie[5] tool is designed to help prevent online plagiarism. All content submitted to this system is checked with around 1 billion web pages. This is a paid system, but free trails are available for this system.

### 3.2 Easy Verification Engine
Easy Verification Engine[6] or EVE is said to be a very powerful tool. It is claimed that this system checks every available website on the internet, to detect plagiarism. This is also a paid system, but 15 days trail is available.

### 3.3 Plagirism.org
Plagiarism.org[7] is a system developed by University California at Barkley (UCB). It detects plagiarism quite efficiently, though it has a drawback. This system cannot differentiate between the original text and the quoted text. It is freely available at UCB's website.

### 3.4 PlagiServe
PlagiServe[8] system has a database of over 150,000 papers, student essays etc. through which it detect plagiarism. Moreover, this system also checks website for potential plagiarism.

### 3.5 Turnitin
Turnitin[9] system is considered to be the best available resource to detect online plagiarism. If original text is available on the internet, it can be detected. This is a paid system, but free trails are available.

### 3.6 Google
Google[10], is a search engine and is not designed to detect plagiarism, but its advance search can help detect plagiarism by extracting key phrases from the text and is quite effective. This is freely available (without any registration).

Various researchers have also studied the problem of plagiarism and have come out with novel approaches. Some of these approaches are ready to be used and some are still be developed. For instance, Eissen and Stein [11][12] have explored the problem and have developed the system where no existing plagiarized corpus is required. Collberg et al[13][14] have developed a system for detecting self plagiarism. Saber and Al-Jafer[15] have developed a system which uses search trees for plagiarism detection in computer software. As most of the programmers copy code online in their projects. This is also a considered as plagiarism.

Attempts have also been made to understand why plagiarism is done. Weinstein & Dobkin[16] discuss how the advent of internet has made act plagiarism very easy. Peirce and Allshouse [17] have studied students perception and use of unfair means to excel among their peer group. Klein [18] provides a comprehensive review of all such studies undertaken by different researchers all around the world.

Since long a need was being felt to develop evaluation measures for various plagiarism detection tools, as there were no sound evaluation measures. Researchers either applied approaches that are being using in evaluation of information retrieval systems or machine translation systems or they simply applied statistical tests. For the first time Potthast el. al. [19] have come out with measures of evaluating plagiarism systems.

## 4. PROPOSED WORK

This project, started as student evaluation system, where student assignments were checked for plagiarism. Yet, it can further be extended for checking research papers submitted in the conferences/journals.

We have developed a document analysis system which performs the entire process in two phases. First, the parser generates parsed sentences from a input files. Second, the detection model checks the parsed files for similarity.

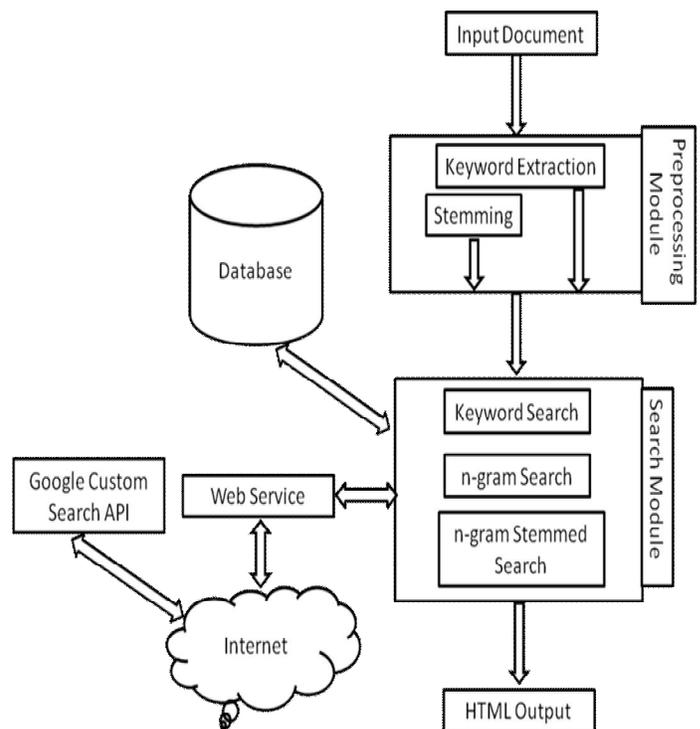

Figure 1. Architecture of Plagiarism Detection System

As shown in figure 1, our model has a web client where the user will input the text document to verify its originality. Originality Checker will execute checking module and finally, the output will be shown. Then we have a web server where the document will be imported for checking.

Here extracted keywords and chucked words are generated and are stored in a database where indexing is performed. Indexed chunks of the sentence are fed to the Google Search Engines using Google's search API available as web service.

Complete document is searched sentence by sentence. First a at the beginning of each sentence, a word in the sentence is selected, then one word to its right and one to its left are combined and are searched. If it is found on some text then to be more specific the phrase is extended by add one word to its left and right and the string is again searched. This process of increasing the word is continued until the search fails or three words to left and right are found. If the search terminates with success then it is stored and next word is searched in the same order. Once a sentence is searched its cumulative score is recorded. The process continues till the end of the sentence. This is the plain search module and at times does not provide very effective results. So, to establish confidence in the process, each word of every sentence is stemmed and is searched in the same process as mentioned above. Web links are stored for each chunked sentence and further are sent to the manipulation module for analysis where plagiarism percentage is computed and a final report is generated to the user. The report is generated page by page, sentence by sentence. Where each copied sentence is marked and a corresponding web link is provided with the line no. and no. of words copied from there.

## 5. CONCLUSION

Many methods have been proposed to detect and stop plagiarism. But, still there are many questions which are to be answered. Natural Language Processing has greater possibilities of providing a sound and concrete mechanism which is capable of detecting plagiarism in any document. In this paper we have tried to show its advantage. We have developed a system based on principles of chunking and keyword extraction, which detects plagiarism.

As an extension to this work, one can include a multi lingual search feature where a keyword in one language can be searched in two or more different languages on the web, as there lay a possibility of cross lingual plagiarism. One can always take some text in one language and translate it to some other language and produce it as his own. We are in a process of developing a Cross Lingual Information Retrieval Tool which can be incorporated in the existing system and would deal with this problem.

Another future work is to develop a sizable collection of plagiarized documents. While testing we faced this problem, we did not had any plagiarized document corpus to test on, so we decided to create some of our own and use most of the student assignments available to us.

Another improvement in our model is analysis of noise in input. This will further improve the performance of the system which would in turn be able to detect documents more accurately.